\documentclass{svmult}
\usepackage{epsfig,graphicx} % figures
\usepackage[bottom]{footmisc}% places footnotes at page bottom
\usepackage{natbib,url}      %RR additions
\usepackage{bigtabular}
\bibpunct{(}{)}{;}{a}{}{,}   %RR reference punctuation
\def\cite#1{\citealp{#1}}    %RR restore old astroncite \cite command
\def\authorindex#1{}  %RA to be redefined by editor at insertion into book

\begin{document}

%%\setcounter{page}{1}
%RA to insert and reset to actual page number for your Astro-PH upload
%RR file: rr-assp-defs.tex = extra ASSP definitions by Rob Rutten
%RR last: Dec 25 2008 
%RR note: ?? problem    %RR Rob-to-Rob   

%RR ## to be adapted when the volume number is known
\def\thisvolume{these proceedings}

%RR journal abbreviations
%%%%%%%%%%%%%%%%%%%%%%%%%
\def\aj{{AJ}}			
\def\araa{{ARA\&A}}		
\def\apj{{ApJ}}			
\def\apjl{{ApJ}}		
\def\apjs{{ApJS}}		
\def\ao{{Appl.\ Optics}} 
\def\apss{{Ap\&SS}}		
\def\aap{{A\&A}}		
\def\aapr{{A\&A~Rev.}}		
\def\aaps{{A\&AS}}		
\def\an{{Astron.\ Nachrichten}}
\def\aspcs{{ASP Conf.\ Ser.}}
\def\azh{{AZh}}			
\def\baas{{BAAS}}		
\def\jrasc{{JRASC}}		
\def\memras{{MmRAS}}		
\def\mnras{{MNRAS}}
\def\nat{{Nat}}		
\def\pra{{Phys.\ Rev.\ A}} 
\def\prb{{Phys.\ Rev.\ B}}		
\def\prc{{Phys.\ Rev.\ C}}		
\def\prd{{Phys.\ Rev.\ D}}		
\def\prl{{Phys.\ Rev.\ Lett}}	
\def\pasp{{PASP}}
\def\pasj{{PASJ}}		
\def\qjras{{QJRAS}}
\def\science{{Sci}}		
\def\skytel{{S\&T}}		
\def\solphys{{Solar\ Phys.}} 
\def\sovast{{Soviet\ Ast.}}  
\def\ssr{{Space\ Sci.\ Rev.}}
\def\svassp{{Astrophys.\ Space Science Proc.}}
\def\zap{{ZAp}}			
\let\astap=\aap
\let\apjlett=\apjl
\let\apjsupp=\apjs

%RR astronomy and math commands copied from ASP
%%%%%%%%%%%%%%%%%%%%%%%%%%%%%%%%%%%%%%%%%%%%%%%
\def\ion#1#2{{\rm #1}\,{\uppercase{#2}}}  %RR ~>\, \sc > uc 
\def\deg{\hbox{$^\circ$}}
\def\sun{\hbox{$\odot$}}
\def\earth{\hbox{$\oplus$}}
\def\la{\mathrel{\hbox{\rlap{\hbox{\lower4pt\hbox{$\sim$}}}\hbox{$<$}}}}
\def\ga{\mathrel{\hbox{\rlap{\hbox{\lower4pt\hbox{$\sim$}}}\hbox{$>$}}}}
\def\sq{\hbox{\rlap{$\sqcap$}$\sqcup$}}
\def\arcmin{\hbox{$^\prime$}}
\def\arcsec{\hbox{$^{\prime\prime}$}}
\def\fd{\hbox{$.\!\!^{\rm d}$}}
\def\fh{\hbox{$.\!\!^{\rm h}$}}
\def\fm{\hbox{$.\!\!^{\rm m}$}}
\def\fs{\hbox{$.\!\!^{\rm s}$}}
\def\fdg{\hbox{$.\!\!^\circ$}}
\def\farcm{\hbox{$.\mkern-4mu^\prime$}}
\def\farcs{\hbox{$.\!\!^{\prime\prime}$}}
\def\fp{\hbox{$.\!\!^{\scriptscriptstyle\rm p}$}}
\def\micron{\hbox{$\mu$m}}
\def\onehalf{\hbox{$\,^1\!/_2$}}	
\def\onethird{\hbox{$\,^1\!/_3$}}
\def\twothirds{\hbox{$\,^2\!/_3$}}
\def\onequarter{\hbox{$\,^1\!/_4$}}
\def\threequarters{\hbox{$\,^3\!/_4$}}
\def\ubv{\hbox{$U\!BV$}}		
\def\ubvr{\hbox{$U\!BV\!R$}}		
\def\ubvri{\hbox{$U\!BV\!RI$}}		
\def\ubvrij{\hbox{$U\!BV\!RI\!J$}}		
\def\ubvrijh{\hbox{$U\!BV\!RI\!J\!H$}}		
\def\ubvrijhk{\hbox{$U\!BV\!RI\!J\!H\!K$}}		
\def\ub{\hbox{$U\!-\!B$}}		
\def\bv{\hbox{$B\!-\!V$}}		
\def\vr{\hbox{$V\!-\!R$}}		
\def\ur{\hbox{$U\!-\!R$}}

%%%%%%%%%%%%%%%%%%%%%%%%%%%%%%%%%%%%%%%%%%%%%%%%%%%%%%%%%%%%%%%%%%%%%%%%%%%%
%RR RJR additional commands
%%%%%%%%%%%%%%%%%%%%%%%%%%%%%%%%%%%%%%%%%%%%%%%%%%%%%%%%%%%%%%%%%%%%%%%%%%%%

%RR -- non-bullet item marker in itemize list 
\def\labelitemi{{\bf --}}  

%RR -- latin abbreviations
\def\rmit#1{{\it #1}}              %% italics (RR style, Kluwer)
\def\rmit#1{{\rm #1}}              %% redefine for ASP, A&A, ApJ, Springer??
\def\etal{\rmit{et al.}}           %% use \etal\ for space behind it        
\def\etc{\rmit{etc.}}           
\def\ie{\rmit{i.e.,}}              %% , required (Webster 1681)
\def\eg{\rmit{e.g.,}}              %% , required (Webster 1681)
\def\cf{cf.}                       %% no Latin, always Roman (Webster 1686)
\def\viz{\rmit{viz.}}
\def\vs{\rmit{vs.}}

%RR -- mathematical
\def\rot{\hbox{\rm rot}}
\def\div{\hbox{\rm div}}
\def\lesssim{\mathrel{\hbox{\rlap{\hbox{\lower4pt\hbox{$\sim$}}}\hbox{$<$}}}}
\def\gtrsim{\mathrel{\hbox{\rlap{\hbox{\lower4pt\hbox{$\sim$}}}\hbox{$>$}}}}
\def\dif{\: {\rm d}}                       %% differential d with space
\def\ep{\:{\rm e}^}                        %% e^ with space and roman e
\def\dash{\hbox{$\,-\,$}}                  %% math-like hyphen
\def\is{\!=\!}                             %% = in text for tighter spacing

%RR --stellar stuff
\def\starname#1#2{${#1}$\,{\rm {#2}}}  %% \starname{\alpha}{Cen~A} 
\def\Teff{\hbox{$T_{\rm eff}$}}   

%RR -- units (in addition to the ASP ones above)
\def\kms{\hbox{km$\;$s$^{-1}$}}
\def\Mxcm{\hbox{Mx\,cm$^{-2}$}}    %% no 2, damn tex

%RR -- magnetic field 
\def\Bapp{\hbox{$B_{\rm app}$}}    %% apparent flux density, Lites convention

%RR -- oscillations
\def\komega{($k, \omega$)}                 %% k - omega 
\def\kf{($k_h,f$)}                         %% f - k_h
\def\VminI{\hbox{$V\!\!-\!\!I$}}           %% V-I
\def\IminI{\hbox{$I\!\!-\!\!I$}}           %% I-I
\def\VminV{\hbox{$V\!\!-\!\!V$}}           %% V-V
\def\Xt{\hbox{$X\!\!-\!t$}}                %% X-t

%RR -- atomic levels
%%      use:    \level 3s3p 3Pe
%%              \level 3s$^2$ {1,3}P{e,o}
%%              \level {} 3Ge
\def\level #1 #2#3#4{$#1 \: ^{#2} \mbox{#3} ^{#4}$}   

%RR -- some spectral species
\def\specchar#1{\uppercase{#1}}    %% to be redefined for A&A = \sc
\def\AlI{\mbox{Al\,\specchar{i}}}  %% use \AlI\ for space behind it
\def\BI{\mbox{B\,\specchar{i}}} 
\def\BII{\mbox{B\,\specchar{ii}}}  
\def\BaI{\mbox{Ba\,\specchar{i}}}  
\def\BaII{\mbox{Ba\,\specchar{ii}}} 
\def\CI{\mbox{C\,\specchar{i}}} 
\def\CII{\mbox{C\,\specchar{ii}}} 
\def\CIII{\mbox{C\,\specchar{iii}}} 
\def\CIV{\mbox{C\,\specchar{iv}}} 
\def\CaI{\mbox{Ca\,\specchar{i}}} 
\def\CaII{\mbox{Ca\,\specchar{ii}}} 
\def\CaIII{\mbox{Ca\,\specchar{iii}}} 
\def\CoI{\mbox{Co\,\specchar{i}}} 
\def\CrI{\mbox{Cr\,\specchar{i}}} 
\def\CriI{\mbox{Cr\,\specchar{ii}}} 
\def\CsI{\mbox{Cs\,\specchar{i}}} 
\def\CsII{\mbox{Cs\,\specchar{ii}}} 
\def\CuI{\mbox{Cu\,\specchar{i}}} 
\def\FeI{\mbox{Fe\,\specchar{i}}} 
\def\FeII{\mbox{Fe\,\specchar{ii}}} 
\def\FeIX{\mbox{Fe\,\specchar{ix}}}
\def\FeX{\mbox{Fe\,\specchar{x}}}
\def\FeXVI{\mbox{Fe\,\specchar{xvi}}}
\def\FrI{\mbox{Fr\,\specchar{i}}}
\def\HI{\mbox{H\,\specchar{i}}} 
\def\HII{\mbox{H\,\specchar{ii}}} 
\def\Hmin{\hbox{\rmH$^{^{_{\scriptstyle -}}}$}}      %% H^min, elegant
\def\Hemin{\hbox{{\rm He}$^{^{_{\scriptstyle -}}}$}} %% He^min, idem
\def\HeI{\mbox{He\,\specchar{i}}} 
\def\HeII{\mbox{He\,\specchar{ii}}} 
\def\HeIII{\mbox{He\,\specchar{iii}}} 
\def\KI{\mbox{K\,\specchar{i}}} 
\def\KII{\mbox{K\,\specchar{ii}}} 
\def\KIII{\mbox{K\,\specchar{iii}}} 
\def\LiI{\mbox{Li\,\specchar{i}}} 
\def\LiII{\mbox{Li\,\specchar{ii}}} 
\def\LiIII{\mbox{Li\,\specchar{iii}}} 
\def\MgI{\mbox{Mg\,\specchar{i}}} 
\def\MgII{\mbox{Mg\,\specchar{ii}}} 
\def\MgIII{\mbox{Mg\,\specchar{iii}}} 
\def\MnI{\mbox{Mn\,\specchar{i}}} 
\def\NI{\mbox{N\,\specchar{i}}}
\def\NaI{\mbox{Na\,\specchar{i}}}
\def\NaII{\mbox{Na\,\specchar{ii}}}
\def\NaIII{\mbox{Na\,\specchar{iii}}} 
\def\NiI{\mbox{Ni\,\specchar{i}}} 
\def\NiII{\mbox{Ni\,\specchar{ii}}}
\def\NiIII{\mbox{Ni\,\specchar{iii}}} 
\def\OI{\mbox{O\,\specchar{i}}} 
\def\OVI{\mbox{O\,\specchar{vi}}}
\def\RbI{\mbox{Rb\,\specchar{i}}} 
\def\SII{\mbox{S\,\specchar{ii}}} 
\def\SiI{\mbox{Si\,\specchar{i}}} 
\def\SiII{\mbox{Si\,\specchar{ii}}} 
\def\SrI{\mbox{Sr\,\specchar{i}}}
\def\SrII{\mbox{Sr\,\specchar{ii}}}
\def\TiI{\mbox{Ti\,\specchar{i}}} 
\def\TiII{\mbox{Ti\,\specchar{ii}}} 
\def\TiIII{\mbox{Ti\,\specchar{iii}}} 
\def\TiIV{\mbox{Ti\,\specchar{iv}}} 
\def\VI{\mbox{V\,\specchar{i}}} 
\def\HtwoO{\mbox{H$_2$O}}        %% H2O %RR TeX doesn't accept numbers alas
\def\Otwo{\mbox{O$_2$}}          %% O2

%RR -- hydrogen spectrum features
\def\Halpha{\mbox{H\hspace{0.1ex}$\alpha$}} %% \Halpha\ for space behind it
\def\Ha{\mbox{H\hspace{0.2ex}$\alpha$}}
\def\Hbeta{\mbox{H\hspace{0.2ex}$\beta$}}
\def\Hgamma{\mbox{H\hspace{0.2ex}$\gamma$}}
\def\Hdelta{\mbox{H\hspace{0.2ex}$\delta$}}
\def\Hepsilon{\mbox{H\hspace{0.2ex}$\epsilon$}}
\def\Hzeta{\mbox{H\hspace{0.2ex}$\zeta$}}
\def\Lyalpha{\mbox{Ly$\hspace{0.2ex}\alpha$}}
\def\Lybeta{\mbox{Ly$\hspace{0.2ex}\beta$}}
\def\Lygamma{\mbox{Ly$\hspace{0.2ex}\gamma$}}
\def\Lycont{\mbox{Ly\hspace{0.2ex}{\small cont}}}
\def\Baalpha{\mbox{Ba$\hspace{0.2ex}\alpha$}}
\def\Babeta{\mbox{Ba$\hspace{0.2ex}\beta$}}
\def\Bacont{\mbox{Ba\hspace{0.2ex}{\small cont}}}
\def\Paalpha{\mbox{Pa$\hspace{0.2ex}\alpha$}}
\def\Bralpha{\mbox{Br$\hspace{0.2ex}\alpha$}}

%RR -- Na D
\def\NaD{\mbox{Na\,\specchar{i}\,D}}    %% use \NaD\ for space behind it
\def\NaDone{\mbox{Na\,\specchar{i}\,\,D$_1$}}
\def\NaDtwo{\mbox{Na\,\specchar{i}\,\,D$_2$}}
\def\NaID{\mbox{Na\,\specchar{i}\,\,D}}
\def\NaIDone{\mbox{Na\,\specchar{i}\,\,D$_1$}}
\def\NaIDtwo{\mbox{Na\,\specchar{i}\,\,D$_2$}}
\def\Done{\mbox{D$_1$}}
\def\Dtwo{\mbox{D$_2$}}

%RR -- Mg b 
\def\Mgbone{\mbox{Mg\,\specchar{i}\,b$_1$}}
\def\Mgbtwo{\mbox{Mg\,\specchar{i}\,b$_2$}}
\def\Mgbthree{\mbox{Mg\,\specchar{i}\,b$_3$}}
\def\MgIb{\mbox{Mg\,\specchar{i}\,b}}
\def\MgIbone{\mbox{Mg\,\specchar{i}\,b$_1$}}
\def\MgIbtwo{\mbox{Mg\,\specchar{i}\,b$_2$}}
\def\MgIbthree{\mbox{Mg\,\specchar{i}\,b$_3$}}

%RR -- Ca II H & K 
\def\CaIIK{\mbox{Ca\,\specchar{ii}\,K}}       %% use \CaIIK\ for space
\def\CaIIH{\mbox{Ca\,\specchar{ii}\,H}}
\def\CaIIHK{\mbox{Ca\,\specchar{ii}\,H\,\&\,K}}
\def\HK{\mbox{H\,\&\,K}}
\def\Kthree{\mbox{K$_3$}}      %% numbers not permitted, alas
\def\Hthree{\mbox{H$_3$}}
\def\Ktwo{\mbox{K$_2$}}
\def\Htwo{\mbox{H$_2$}}
\def\Kone{\mbox{K$_1$}}     
\def\Hone{\mbox{H$_1$}}     
\def\KtwoV{\mbox{K$_{2V}$}}
\def\KtwoR{\mbox{K$_{2R}$}}
\def\KoneV{\mbox{K$_{1V}$}}
\def\KoneR{\mbox{K$_{1R}$}}
\def\HtwoV{\mbox{H$_{2V}$}}
\def\HtwoR{\mbox{H$_{2R}$}}
\def\HoneV{\mbox{H$_{1V}$}}
\def\HoneR{\mbox{H$_{1R}$}}

%RR -- Mg II h & k 
\def\hk{\mbox{h\,\&\,k}}
\def\kthree{\mbox{k$_3$}}    
\def\hthree{\mbox{h$_3$}}
\def\ktwo{\mbox{k$_2$}}
\def\htwo{\mbox{h$_2$}}
\def\kone{\mbox{k$_1$}}     
\def\hone{\mbox{h$_1$}}     
\def\ktwoV{\mbox{k$_{2V}$}}
\def\ktwoR{\mbox{k$_{2R}$}}
\def\koneV{\mbox{k$_{1V}$}}
\def\koneR{\mbox{k$_{1R}$}}
\def\htwoV{\mbox{h$_{2V}$}}
\def\htwoR{\mbox{h$_{2R}$}}
\def\honeV{\mbox{h$_{1V}$}}
\def\honeR{\mbox{h$_{1R}$}}

%RA use these as needed but don't change nor add!
%RA   if you really need private \def's then define them here
%RA   so that we can inspect them and put into rr-assp-defs
%RA   if really useful - and not upsetting any other paper

\title*{Helicity at Photospheric and Chromospheric Heights}
%RA capitalize the nouns, try to fit on one line

\author{S. K. Tiwari\inst{1},
         P. Venkatakrishnan\inst{1}
         \and
         K. Sankarasubramanian\inst{2}}
%RA full initials but no first names; spaces between initials please.
\authorindex{Tiwari, S. K.}
\authorindex{Venkatakrishnan, P.}
\authorindex{Sankarasubramanian, K.}

\institute{Udaipur Solar Observatory (Physical Research Laboratory),
    Udaipur, India
           \and
           Space Astron.\ \& Instrument.\ Div.,  ISRO Satellite Center, Bangalore, India }
%RA no full postal addresses, just brief affiliation; no email addresses
\maketitle
\setcounter{footnote}{0}  %RR Springer forgot this one (and much more)

%%%%%%%%%%%%%%%%%%%%%%%%%%%%%%%%%%%%%%%%%%%%%%%%%%%%%%%%%%%%%%%%%%%%%%%%%%%%
\begin{abstract}
   In the solar atmosphere the twist parameter $\alpha$ has the
   same sign as magnetic helicity. It has been observed using
   photospheric vector magnetograms that negative/positive helicity is
   dominant in the northern/southern hemisphere of the
   Sun. Chromospheric features show dextral/sinistral dominance in the
   northern/southern hemisphere and sigmoids observed in X-rays also
   have a dominant sense of reverse-S/forward-S in the
   northern/southern hemisphere. It is of interest whether individual
   features have one-to-one correspondence in terms of helicity at
   different atmospheric heights.  We use UBF \Halpha\ images from the
   Dunn Solar Telescope (DST) and other \Halpha\ data from Udaipur
   Solar Observatory and Big Bear Solar Observatory. Near-simultaneous
   vector magnetograms from the DST are used to establish one-to-one
   correspondence of helicity at photospheric and chromospheric
   heights. We plan to extend this investigation with more data
   including coronal intensities.
\end{abstract}
%%%%%%%%%%%%%%%%%%%%%%%%%%%%%%%%%%%%%%%%%%%%%%%%%%%%%%%%%%%%%%%%%%%%%%%%%%%%
%RA no keywords please

%%%%%%%%%%%%%%%%%%%%%%%%%%%%%%%%%%%%%%%%%%%%%%%%%%%%%%%%%%%%%%%%%%%%%%%%%%%%
\section{Introduction}      \label{rutten-sec:introduction}
%%%%%%%%%%%%%%%%%%%%%%%%%%%%%%%%%%%%%%%%%%%%%%%%%%%%%%%%%%%%%%%%%%%%%%%%%%%%
Helicity is a physical quantity that measures the degree of linkage
and twistedness in the field (\cite{1984JFM...147..133B}).  It is derived from a
volume integral over the scalar product of the magnetic field
\textbf{B} and its vector potential \textbf{A}. Direct calculation of
helicity is not possible due to the non-uniqueness of the vector
potential \textbf{A} and the limited availability of data sampling
different layers of the solar atmosphere.  The force-free parameter
$\alpha$ estimates one component of helicity, \ie\
twist, the other component being writhe which can not be derived from
the available data. This $\alpha$ is a measure of degree of twist per unit
axial length.  It has the same sign as magnetic helicity
(\cite{2008ApJ...677..719P},
\cite{2008JApA...29...49P}).  It is now well known that
negative/positive helicity dominates in the northern/southern hemisphere.
For active regions the hemispheric helicity rule holds in the
photosphere, see \citet{2005PASJ...57..481H} and references
therein. Similarly for the chromospheric and coronal helicity
rules, see \citet{2005SoPh..228...97B} and references therein, and
\citet{2001ApJ...549L.261P} and references therein. The
topology of chromospheric and coronal features decide the sign of
the associated helicity.  Chirality is the term used for the sign of
the helicity in these features.  Thus, helicity is a physical
measure of chirality. The chirality of active region features shows
correspondence with the sign of the helicity in the associated lower/upper
atmospheric features. For example, the chirality of X-ray features
with S (inverse-S) shapes are associated with sinistral (dextral)
filaments (\cite{2003AdSpR..32.1883M}, \cite{2003AdSpR..32.1895R}).
\citet{2000ApJ...540L.115C} reported for a few cases
that active filaments showing dextral/sinistral chirality are related
with negative/positive magnetic helicity.
\citet{2001ApJ...549L.261P} demonstrated correspondence
between photospheric and coronal chirality for a few active regions.
However, this needs to be confirmed.  We have reported similar
helicity signs at photospheric, chromospheric, and coronal heights for
a few active regions (\cite{2008tdad.conf..329T}).

Comparison between magnetic helicity signs at different heights in the
solar atmosphere may be a useful tool to predict solar eruptions
leading to interplanetary events. Also, it may help to constrain
modeling chromospheric and coronal features taking the photosphere as
boundary condition.  However, the data required to do this are not
directly available and are often non-conclusive.  Vector magnetic
fields are not available as routinely as is necessary to derive
photospheric twist values.  Chromospheric \Halpha\ images may be
available most of the time by combining data from different
telescopes, but are not always conclusive due to lack of angular
resolution.  Analysis of coronal loop observations is required to
determine coronal helicity signs, but these are also not available
routinely.  Above all, it is hard to find data taken simultaneously at
different heights in the solar atmosphere.  In this work we combine
photospheric and chromospheric data from multiple solar observatories
and telescopes.  They were often not taken at precisely the same time.
We therefore assume that the sign of the magnetic helicity does not
change within a few hours.

%%%%%%%%%%%%%%%%%%%%%%%%%%%%%%%%%%%%%%%%%%%%%%%%%%%%%%%%%%%%%%%%%%%%%%%%%%%%
\section{Sign of Magnetic Helicity }                 %  \label{rutten-sec:evidence}
%%%%%%%%%%%%%%%%%%%%%%%%%%%%%%%%%%%%%%%%%%%%%%%%%%%%%%%%%%%%%%%%%%%%%%%%%%%%
The sign of helicity in the photosphere is usually found from the
force-free parameter $\alpha$,
\eg\  $\alpha_{\rm best}$ (\cite{1995ApJ...440L.109P}), averaged $\alpha$, 
\eg\ $<\alpha_{z}>$ = $<\bf{J_{z}} / \bf{B_{z}}>$
 (\cite{1994ApJ...425L.117P}) with current density $\bf{J_{z}} =
 \bf\nabla\times \bf{B_{z}}$ , where $\bf{B_{z}}$ is the vertical
 component of the magnetic field.  Some authors have used the current
 helicity density $H_{c }= \bf{B_{z}} \cdot \bf{J_{z}} $
 (\cite{1998ApJ...496L..43B}; \cite{2005PASJ...57..481H}).  A good
 correlation was found between $\alpha_{\rm best}$ and
 $\langle\alpha_{z}\rangle$ by \citet{2004ApJ...606..565B} and
 \citet{1996ApJ...462..547L}. The force-free parameter $\alpha$ has
 the same sign as magnetic helicity (\cite{2008ApJ...677..719P}).
 Also, the current helicity (which is not a conserved quantity like
 magnetic helicity) has the same sign as that of magnetic helicity
 (\cite{1990SoPh..125..219S}; \cite{1999SoPh..189...25H};
 \cite{2008JApA...29...49P}; \cite{2008SoPh..248...17S}).  In this
 study, we use the sign of the global $\alpha$ parameter as the sign
 of magnetic helicity, giving the twist present in the active region.

%%%%%%%%%%%%%%%%%%%%%%%%%%%%%%%%%%%%%%%%%%%%%%%%%%%%%%%%%%%%%%%%%%%%%%%%%%%%
%\section{Chromospheric sign of magnetic helicity }
%%%%%%%%%%%%%%%%%%%%%%%%%%%%%%%%%%%%%%%%%%%%%%%%%%%%%%%%%%%%%%%%%%%%%%%%%%%%
Numerical measurement of the sign of the chromospheric magnetic
helicity is not possible due to non-availability of vector magnetic
field observations at these heights.  However, the twist present in
morphological intensity features were reported already long ago
(\cite{1925PASP...37..268H}; \cite{1941ApJ....93...24R}) to tend to
follow the hemispheric helicity rule, independent of the
solar cycle. Later, many researchers studied the chirality of
different chromospheric features such as filaments, fibrils, filament
channels etc. \citep{1998ASPC..150..419M,2003AdSpR..32.1883M}. We use the
chirality of these chromospheric features, mostly whirls observed in
\Halpha, to derive its association with the photospheric
sign of magnetic helicity.

%RT you can do that on many DOT movies too

%%%%%%%%%%%%%%%%%%%%%%%%%%%%%%%%%%%%%%%%%%%%%%%%%%%%%%%%%%%%%%%%%%%%%%%%%%%%
\section{Data Sets Used }
%%%%%%%%%%%%%%%%%%%%%%%%%%%%%%%%%%%%%%%%%%%%%%%%%%%%%%%%%%%%%%%%%%%%%%%%%%%%
Apart from a few data sets, most are obtained from different solar
observatories and telescopes due to the unavailability of all required
data from a at the same place.  The vector magnetic field data were obtained
from the Advanced Stokes Polarimeter (ASP: \cite{1992SPIE.1746...22E})
as well as the Diffraction Limited Spectro-polarimeter (DLSP:
\citep{2004SPIE.5171..207S,2006ASPC..358..201S}) both at the DST.
Near-simultaneous
\Halpha\ images from the Universal Bi-refringent Filter (UBF)
at the DST are used whenever obtained along with the ASP and DLSP.
For the vector field observations which do not have corresponding UBF
data, \Halpha\ images from Udaipur Solar Observatory (USO) and Big
Bear Solar Observatory (BBSO) were used.  We made sure that in these
cases the \Halpha\ images were collected within less than a
day. Standard and well-established processing was done to derive
vector fields.  The procedure is described in the references given
above.

\begin{figure}
\centerline
 {\epsfxsize=2.70in\epsfysize=2.70in\epsfbox{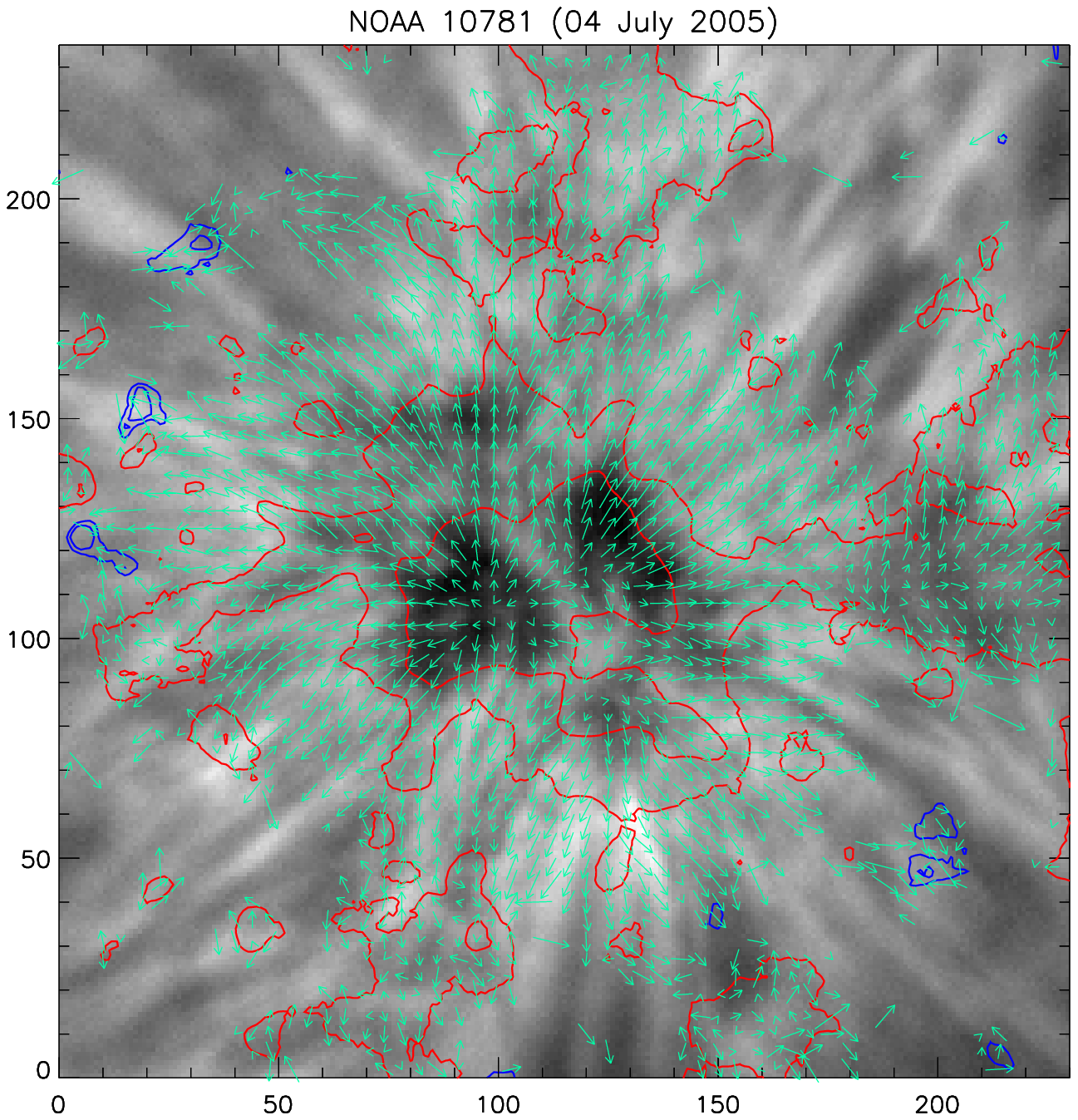}
 \epsfxsize=2.70in\epsfysize=2.70in\epsfbox{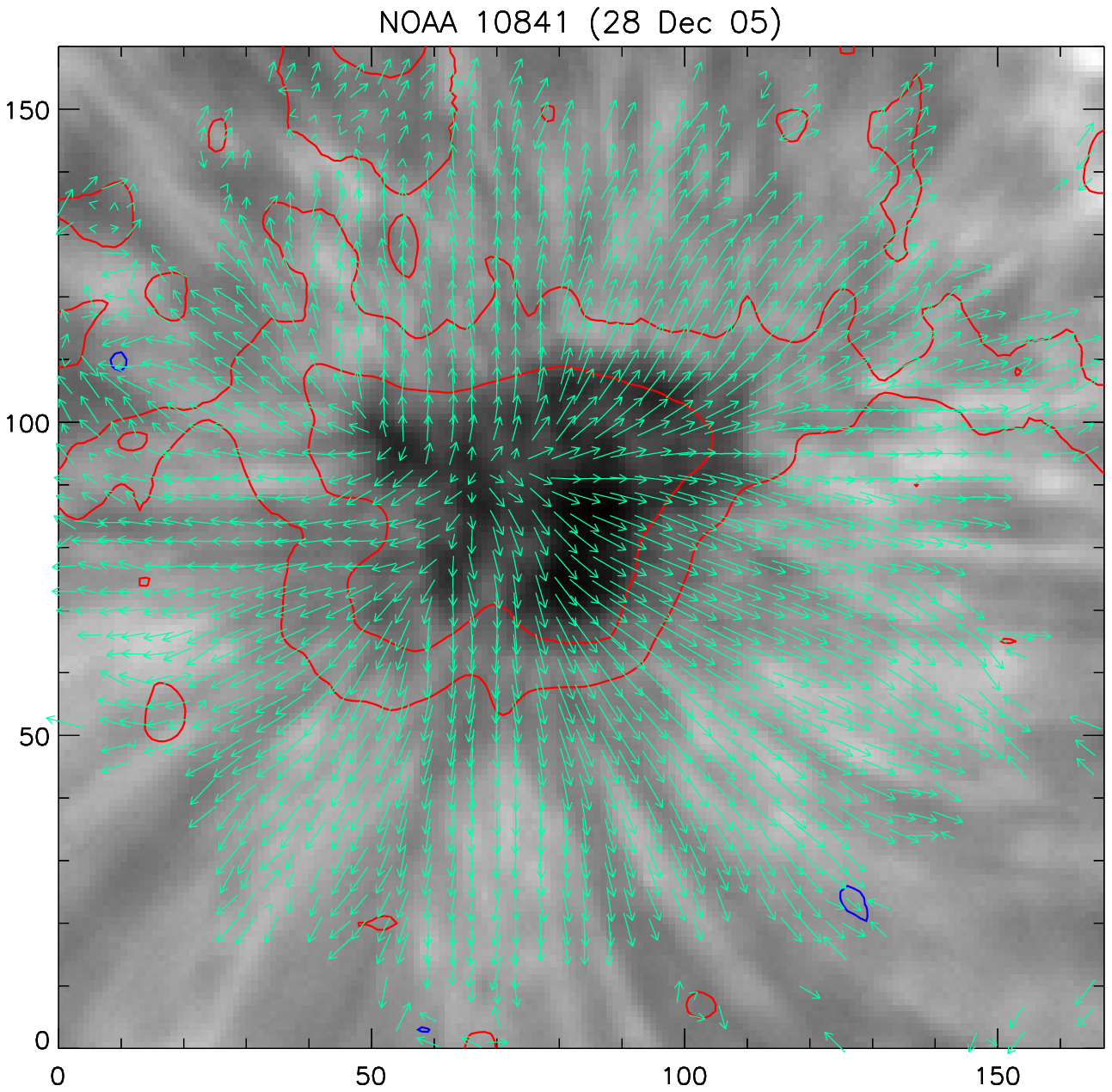}}
\caption {Two examples of the chromospheric sunspots, with
the photospheric transverse vectors of the same field of view
over-plotted on them. The axis
divisions are pixel numbers. The contours and vectors show
longitudinal and transverse magnetic fields, respectively.}
\label{tiwari-fig:fig1}
\end{figure}

\begin{table}
\caption{Sign of helicity at the photospheric and the chromospheric
level \label{tbl-1}}
\centering
\begin{tabular}{c c c c}
\hline     %%for horizontal space
Observation date & AR NOAA number & Photospheric helicity & Chromospheric chirality \\
\hline
06 Feb 2007 & NOAA 10941 & Negative & Dextral \\	
09 Jan 2007 & NOAA 10935 & Negative & Dextral \\	
11 Dec 2006 & NOAA 10930 & Negative & Dextral \\
28 Dec 2005 & NOAA 10841 & Positive & Sinistral{$^{\sharp}$} \\	
22 Dec 2005 & NOAA 10838 & Positive & Sinistral	\\
27 Aug 2005 & NOAA 10804 & Positive & Sinistral	\\
28 Aug 2005 & NOAA 10803 & Positive & Sinistral	\\
26 Aug 2005 & NOAA 10800 & Positive & Sinistral	\\
04 July 2005 & NOAA 10781 & Positive & Sinistral{$^{\sharp}$} \\	
17 Apr 2005 & NOAA 10752 & Positive & Sinistral	\\
09 Apr 2003 & NOAA 10330 & Positive & Sinistral{$^{\sharp}$}	\\
03 Sep 2001 & NOAA 09601 & Negative & Dextral	\\
01 Sep 2001 & NOAA 09601 & Negative & Dextral{$^{\sharp}$}	\\
30 Aug 2001 & NOAA 09601 & Negative & Dextral	\\
01 Sep 2001 & NOAA 09596 & Positive & Sinistral	\\
30 Aug 2001 & NOAA 09596 & Positive & Sinistral \\	
30 Aug 2001 & NOAA 09591 & Positive & Sinistral	\\
26 Aug 2001 & NOAA 09590 & Negative & Dextral	\\
24 Aug 2001 & NOAA 09590 & Negative & Dextral	\\
24 Aug 2001	& NOAA 09585 & Positive & Sinistral{$^{\sharp}$} \\

\hline
$\sharp$ : \it dominant sense
\end{tabular}
\end{table}

\section{Results and Discussion}
Table~\ref{tbl-1}
shows how the sign of helicity at the photospheric level and of the
chirality in
associated features at chromospheric heights are related with each
other. Figure~\ref{tiwari-fig:fig1} (a) and (b) clearly show that the
\Halpha\ whirls follow the transverse magnetic field vectors
measured at photospheric heights. The positive/negative helicity
derived from the global twist in this sunspot are in the photospheric
vector data is directly associated with the sinistral/dextral
chirality derived from the chromospheric \Halpha\ data.

In this preliminary analysis, we thus conclude that the sign of
helicity (positive/negative) derived from global twist present around
sunspots in the photosphere has one-to-one correspondence with the
(sinistral/dextral) sense of chirality observed in the associated
chromospheric data.  We mostly use the chirality of chromospheric
whirls to derive the chromospheric helicity sign.  It is known
\citep{1998ASPC..150..419M,2003AdSpR..32.1883M} that filaments, filament
channels, etc., have the same sense of chirality as the whirls above
the associated active regions.  The chirality of filaments associated
with an active region can therefore be used to determine the
chromospheric sense of chirality when high resolution
\Halpha\ data are not available.

%%%%%%%%%%%%%%%%%%%%%%%%%%%%%%%%%%%%%%%%%%%%%%%%%%%%%%%%%%%%%%%%%%%%%%%%%%%%
\begin{acknowledgement}
  We thank the conference organisers for a very good meeting and the
  editors for excellent instructions.
\end{acknowledgement}

%%%%%%%%%%%%%%%%%%%%%%%%%%%%%%%%%%%%%%%%%%%%%%%%%%%%%%%%%%%%%%%%%%%%%%%%%%%%
%% References
%%%%%%%%%%%%%%%%%%%%%%%%%%%%%%%%%%%%%%%%%%%%%%%%%%%%%%%%%%%%%%%%%%%%%%%%%%%%
\begin{small}

\bibliographystyle{rr-assp}       %RR hacked from aa.bst
\bibliography{tiwari}

\begin{thebibliography}{24}
\expandafter\ifx\csname natexlab\endcsname\relax\def\natexlab#1{#1}\fi

\bibitem[{{Bao} \& {Zhang}(1998)}]{1998ApJ...496L..43B}
{Bao}, S. {Zhang}, H. 1998, \apjl, 496, L43+

\bibitem[{{Berger} \& {Field}(1984)}]{1984JFM...147..133B}
{Berger}, M.~A. {Field}, G.~B. 1984, Journal of Fluid Mechanics, 147, 133

\bibitem[{{Bernasconi} {et~al.}(2005){Bernasconi}, {Rust}, \&
  {Hakim}}]{2005SoPh..228...97B}
{Bernasconi}, P.~N., {Rust}, D.~M., {Hakim}, D. 2005, \solphys, 228, 97

\bibitem[{{Burnette} {et~al.}(2004){Burnette}, {Canfield}, \&
  {Pevtsov}}]{2004ApJ...606..565B}
{Burnette}, A.~B., {Canfield}, R.~C., {Pevtsov}, A.~A. 2004, \apj, 606, 565

\bibitem[{{Chae}(2000)}]{2000ApJ...540L.115C}
{Chae}, J. 2000, \apjl, 540, L115

\bibitem[{{Elmore} {et~al.}(1992){Elmore}, {Lites}, {Tomczyk}, {Skumanich},
  {Dunn}, {Schuenke}, {Streander}, {Leach}, {Chambellan}, \&
  {Hull}}]{1992SPIE.1746...22E}
{Elmore}, D.~F., {Lites}, B.~W., {Tomczyk}, S., {et~al.} 1992, in Society of
  Photo-Optical Instrumentation Engineers (SPIE) Conference Series, eds. D.~H.
  {Goldstein} \& R.~A. {Chipman}, Society of Photo-Optical Instrumentation
  Engineers (SPIE) Conference Series, 1746, 22

\bibitem[{{Hagino} \& {Sakurai}(2005)}]{2005PASJ...57..481H}
{Hagino}, M. {Sakurai}, T. 2005, \pasj, 57, 481

\bibitem[{{Hagyard} \& {Pevtsov}(1999)}]{1999SoPh..189...25H}
{Hagyard}, M.~J. {Pevtsov}, A.~A. 1999, \solphys, 189, 25

\bibitem[{{Hale}(1925)}]{1925PASP...37..268H}
{Hale}, G.~E. 1925, \pasp, 37, 268

\bibitem[{{Leka} {et~al.}(1996){Leka}, {Canfield}, {McClymont}, \& {van
  Driel-Gesztelyi}}]{1996ApJ...462..547L}
{Leka}, K.~D., {Canfield}, R.~C., {McClymont}, A.~N., {van Driel-Gesztelyi}, L.
  1996, \apj, 462, 547

\bibitem[{{Martin}(1998)}]{1998ASPC..150..419M}
{Martin}, S.~F. 1998, in IAU Colloq. 167: New Perspectives on Solar
  Prominences, eds. D.~F. {Webb}, B.~{Schmieder}, \& D.~M. {Rust}, Astronomical
  Society of the Pacific Conference Series, 150,  419

\bibitem[{{Martin}(2003)}]{2003AdSpR..32.1883M}
{Martin}, S.~F. 2003, Advances in Space Research, 32, 1883

\bibitem[{{Pevtsov}(2008)}]{2008JApA...29...49P}
{Pevtsov}, A.~A. 2008, Journal of Astrophysics and Astronomy, 29, 49

\bibitem[{{Pevtsov} {et~al.}(2001){Pevtsov}, {Canfield}, \&
  {Latushko}}]{2001ApJ...549L.261P}
{Pevtsov}, A.~A., {Canfield}, R.~C., {Latushko}, S.~M. 2001, \apjl, 549, L261

\bibitem[{{Pevtsov} {et~al.}(1994){Pevtsov}, {Canfield}, \&
  {Metcalf}}]{1994ApJ...425L.117P}
{Pevtsov}, A.~A., {Canfield}, R.~C., {Metcalf}, T.~R. 1994, \apjl, 425, L117

\bibitem[{{Pevtsov} {et~al.}(1995){Pevtsov}, {Canfield}, \&
  {Metcalf}}]{1995ApJ...440L.109P}
{Pevtsov}, A.~A., {Canfield}, R.~C., {Metcalf}, T.~R. 1995, \apjl, 440, L109

\bibitem[{{Pevtsov} {et~al.}(2008){Pevtsov}, {Canfield}, {Sakurai}, \&
  {Hagino}}]{2008ApJ...677..719P}
{Pevtsov}, A.~A., {Canfield}, R.~C., {Sakurai}, T., {Hagino}, M. 2008, \apj,
  677, 719

\bibitem[{{Richardson}(1941)}]{1941ApJ....93...24R}
{Richardson}, R.~S. 1941, \apj, 93, 24

\bibitem[{{Rust}(2003)}]{2003AdSpR..32.1895R}
{Rust}, D.~M. 2003, Advances in Space Research, 32, 1895

\bibitem[{{Sankarasubramanian} {et~al.}(2004){Sankarasubramanian}, {Gullixson},
  {Hegwer}, {Rimmele}, {Gregory}, {Spence}, {Fletcher}, {Richards}, {Rousset},
  {Lites}, {Elmore}, {Streander}, \& {Sigwarth}}]{2004SPIE.5171..207S}
{Sankarasubramanian}, K., {Gullixson}, C., {Hegwer}, S., {et~al.} 2004, in
  Society of Photo-Optical Instrumentation Engineers (SPIE) Conference Series,
  eds. S.~{Fineschi} \& M.~A. {Gummin}, Society of Photo-Optical
  Instrumentation Engineers (SPIE) Conference Series, 5171, 207

\bibitem[{{Sankarasubramanian} {et~al.}(2006){Sankarasubramanian}, {Lites},
  {Gullixson}, {Elmore}, {Hegwer}, {Streander}, {Rimmele}, {Fletcher},
  {Gregory}, \& {Sigwarth}}]{2006ASPC..358..201S}
{Sankarasubramanian}, K., {Lites}, B., {Gullixson}, C., {et~al.} 2006, in
  Astronomical Society of the Pacific Conference Series, eds. R.~{Casini} \&
  B.~W. {Lites}, Astronomical Society of the Pacific Conference Series, 358,
  201

\bibitem[{{Seehafer}(1990)}]{1990SoPh..125..219S}
{Seehafer}, N. 1990, \solphys, 125, 219

\bibitem[{{Sokoloff} {et~al.}(2008){Sokoloff}, {Zhang}, {Kuzanyan}, {Obridko},
  {Tomin}, \& {Tutubalin}}]{2008SoPh..248...17S}
{Sokoloff}, D., {Zhang}, H., {Kuzanyan}, K.~M., {et~al.} 2008, \solphys, 248,
  17

\bibitem[{{Tiwari} {et~al.}(2008){Tiwari}, {Joshi}, {Gosain}, \&
  {Venkatakrishnan}}]{2008tdad.conf..329T}
{Tiwari}, S.~K., {Joshi}, J., {Gosain}, S., {Venkatakrishnan}, P. 2008, in
  Turbulence, Dynamos, Accretion Disks, Pulsars and Collective Plasma
  Processes, eds. S.~S. {Hasan}, R.~T. {Gangadhara}, \& V.~{Krishan}, 329

\end{thebibliography}

\end{small}

\end{document}